\documentstyle[12pt]{article}

\def\baselinestretch{1.2}
\catcode`\@=11
\def\section{\@startsection {section}{1}{\z@}{3.ex plus 1ex minus
 .2ex}{2.ex plus .2ex}{\raggedright\large\bf}}
\def\subsection{\@startsection{subsection}{2}{\z@}{2.75ex plus 1ex minus
 .2ex}{1.5ex plus .2ex}{\raggedright\bf}}
\def\appendix{{\newpage\section*{Appendices}}\let\appendix\section%
        {\setcounter{section}{0}
        \gdef\thesection{\Alph{section}}}\section}
\catcode`\@=12
\newskip\humongous \humongous=0pt plus 1000pt minus 1000pt
\def\caja{\mathsurround=0pt}
\def\eqalign#1{\,\vcenter{\openup1\jot \caja
        \ialign{\strut \hfil$\displaystyle{##}$&$
        \displaystyle{{}##}$\hfil\crcr#1\crcr}}\,}
\newif\ifdtup

\def\oldreffmt#1{\rlap{[#1]} \hbox to 2\parindent{}}

\def\figfmt#1{\rlap{Figure {#1}} \hbox to 1in{}}

\def\gtap{\raisebox{-.4ex}{\rlap{$\sim$}} \raisebox{.4ex}{$>$}}

\def\beq{\begin{equation}}
\def\eeq{\end{equation}}
\def\bea{\begin{eqnarray}}

\def\com#1#2{
        \left[#1, #2\right]}
\def\eea{\end{eqnarray}}
\def\ap#1,#2,#3#4{           {\it Ann. Phys. (NY)\/ }{\bf #1} (19#3#4) #2}
\def\apj#1,#2,#3#4{          {\it Astrophys. J.\/ }{\bf #1} (19#3#4) #2}
\def\apjl#1,#2,#3#4{         {\it Astrophys. J. Lett.\/ }{\bf #1} (19#3#4) #2}
\def\app#1,#2,#3#4{          {\it Acta Phys. Polon.\/ }{\bf #1} (19#3#4) #2}
\def\com#1,#2,#3#4{          {\it Comm. Math. Phys.\/ }{\bf #1} (19#3#4) #2}
\def\ib#1,#2,#3#4{           {\it ibid.\/ }{\bf #1} (19#3#4) #2}
\def\nat#1,#2,#3#4{          {\it Nature (London)\/ }{\bf #1} (19#3#4) #2}
\def\np#1,#2,#3#4{           {\it Nucl. Phys.\/ }{\bf B#1} (19#3#4) #2}
\def\npps#1,#2,#3#4{         {\it Nucl. Phys. B (Proc. Suppl.)\/ }{\bf B#1}
                             (19#3#4) #2}
\def\pl#1,#2,#3#4{           {\it Phys. Lett.\/ }{\bf #1B} (19#3#4) #2}
\def\pla#1,#2,#3#4{          {\it Phys. Lett.\/ }{\bf #1A} (19#3#4) #2}
\def\pr#1,#2,#3#4{           {\it Phys. Rev.\/ }{\bf D#1} (19#3#4) #2}
\def\prep#1,#2,#3#4{         {\it Phys. Rep.\/ }{\bf #1} (19#3#4) #2}
\def\prl#1,#2,#3#4{          {\it Phys. Rev. Lett.\/ }{\bf #1} (19#3#4) #2}
\def\pro#1,#2,#3#4{          {\it Prog. Theor. Phys.\/ }{\bf #1} (19#3#4) #2}
\def\rmp#1,#2,#3#4{          {\it Rev. Mod. Phys.\/ }{\bf #1} (19#3#4) #2}
\def\sp#1,#2,#3#4{           {\it Sov. Phys.-Usp.\/ }{\bf #1} (19#3#4) #2}

\def\zp#1,#2,#3#4{           {\it Zeit. fur Physik\/ }{\bf #1} (19#3#4) #2}
\catcode`\@=11
\def\eqnarray{\stepcounter{equation}\let\@currentlabel=\theequation
\global\@eqnswtrue
\global\@eqcnt\z@\tabskip\@centering\let\\=\@eqncr
\gdef\@@fix{}\def\eqno##1{\gdef\@@fix{##1}}%
$$\halign to \displaywidth\bgroup\@eqnsel\hskip\@centering
  $\displaystyle\tabskip\z@{##}$&\global\@eqcnt\@ne
  \hskip 2\arraycolsep \hfil${##}$\hfil
  &\global\@eqcnt\tw@ \hskip 2\arraycolsep $\displaystyle\tabskip\z@{##}$\hfil
   \tabskip\@centering&\llap{##}\tabskip\z@\cr}

\def\@@eqncr{\let\@tempa\relax
    \ifcase\@eqcnt \def\@tempa{& & &}\or \def\@tempa{& &}
      \else \def\@tempa{&}\fi
     \@tempa \if@eqnsw\@eqnnum\stepcounter{equation}\else\@@fix\gdef\@@fix{}\fi
     \global\@eqnswtrue\global\@eqcnt\z@\cr}

\newcount\hour
\newcount\minute
\newtoks\amorpm
\hour=\time\divide\hour by 60
\minute=\time{\multiply\hour by 60 \global\advance\minute by-\hour}
\edef\standardtime{{\ifnum\hour<12 \global\amorpm={am}%
	\else\global\amorpm={pm}\advance\hour by-12 \fi
	\ifnum\hour=0 \hour=12 \fi
	\number\hour:\ifnum\minute<10 0\fi\number\minute\the\amorpm}}
\edef\militarytime{\number\hour:\ifnum\minute<10 0\fi\number\minute}

\def\draftlabel#1{{\@bsphack\if@filesw {\let\thepage\relax
   \xdef\@gtempa{\write\@auxout{\string
      \newlabel{#1}{{\@currentlabel}{\thepage}}}}}\@gtempa
   \if@nobreak \ifvmode\nobreak\fi\fi\fi\@esphack}
        \gdef\@eqnlabel{#1}}
\def\@eqnlabel{}
\def\@vacuum{}
\def\marginnote#1{}
\def\draftmarginnote#1{\marginpar{\raggedright\scriptsize\tt#1}}
\def\draft{
	\pagestyle{plain}
	\overfullrule=2pt
        \oddsidemargin -.5truein
        \def\@oddhead{\sl \phantom{\today\quad\militarytime} \hfil
        \smash{\Large\sl DRAFT} \hfil \today\quad\militarytime}
        \let\@evenhead\@oddhead
        \let\label=\draftlabel
        \let\marginnote=\draftmarginnote
        \def\ps@empty{\let\@mkboth\@gobbletwo
        \def\@oddfoot{\hfil \smash{\Large\sl DRAFT} \hfil}
        \let\@evenfoot\@oddhead}
        \def\@eqnnum{(\theequation)\rlap{\kern\marginparsep\tt\@eqnlabel}%
        \global\let\@eqnlabel\@vacuum}  }

\def\theequation{\thesection.\arabic{equation}}
\@addtoreset{equation}{section}
\@addtoreset{footnote}{section}

\catcode`\@=12

\input epsf
\def\cpsbox{\epsfcheck\cpsbox}
\def\epsfcheck{\ifx\epsfbox\UnDeFiNeD
	\message{(NO epsf.tex, FIGURES WILL BE IGNORED)}
	\gdef\cpsbox##1##2{\vbox to 2in{\hbox to ##1 {\hss} \vss}}
\else\gdef\cpsbox##1##2{
	\setlength{\epsfxsize}{##1}
	\centerline{\epsfbox{##2}}}\fi}
\def\psinsert#1#2#3{
        \begin{figure}
  	\cpsbox{#1 \hsize}{#2}
	\medskip
	\centerline{
	   \vbox{\hsize=#1\hsize \footnotesize \def\baselinestretch{1.} #3 } }
        \end{figure}
}

\def\ie{\hbox{\it i.e.\/}}

\def\lae{\smash{\,\lower .5 ex \hbox{$\,\stackrel<\sim\,$}}}
\def\gae{\smash{\,\lower .5 ex \hbox{$\,\stackrel>\sim\,$}}}

\def\L{{\cal L}}

\def\beq{\begin{equation}}
\def\eeq{\end{equation}}

\def\sutw{${\rm SU}(2)_W$}

\def\KKb{$K^0-\bar K^0$}
\def\DDb{$D^0-\bar D^0$}
\def\Kpinunu{$K^+\longrightarrow\pi^+\nu\bar\nu$}
\def\Kee{$K_L\longrightarrow e^+e^-$}
\def\Kenu{$K\longrightarrow e\nu$}
\def\Kmunu{$K\longrightarrow \mu\nu$}

\begin{document}
\begin{titlepage}
\begin{center}
September 15, 1993\hfill    WIS--93/90/Sept--PH

\vskip 1 cm

{\large \bf  A Comprehensive Study of Leptoquark Bounds}

\vskip 1 cm

Miriam Leurer

\vskip 1 cm

{\em Department of Particle Physics\\
The Weizmann Institute\\
Rehovot 76100\\
ISRAEL}

\end{center}

\vskip 1 cm

\begin{abstract}
We make a comprehensive study of indirect bounds on scalar leptoquarks that
couple chirally and diagonally to the first generation by examining available
data from low energy experiments as well as from high energy $e^+e^-$ and
$p\bar p$ accelerators.

The strongest bounds turn out to arise from low energy data: For leptoquarks
that couple to right--handed quarks, the most stringent bound comes from atomic
parity violation. For leptoquarks that couple to left--handed quarks, there are
two mass regions: At low masses the bounds arise from atomic parity violation
or from universality in leptonic $\pi$ decays. At masses above a few hundred
GeV's the dominant bounds come from the FCNC processes that are unavoidable in
these leptoquarks: The FCNC bound of the up sector, that arises from \DDb{}
mixing, combines with the FCNC bounds from the down sector, that arise from
rare $K$ decays and \KKb{} mixing, to a bound on the flavour {\it conserving}
coupling to the first generation.

The bounds restrict leptoquarks that couple with electromagnetic strength to
lie above $600$~GeV or $630$~GeV for leptoquarks that couple to RH quarks, and
above $1040$~GeV, $440$~GeV, and $750$~GeV for the \sutw{} scalar, doublet and
triplet leptoquarks that couple to LH quarks. These bounds are considerably
stronger than the first results from the direct searches at HERA. Our bounds
also already exclude large regions in the parameter space that could be
examined
by various methods proposed for indirect leptoquark searches.

\end{abstract}

\end{titlepage}
\newpage
\section{Introduction}

The original motivation for this research was to compare the oncoming results
from the direct leptoquark search at HERA \cite{HERA} with indirect bounds that
are available from various low energy experiments and from $e^+e^-$ and $p \bar
p$ colliders. A previous study \cite{Buch1} of such indirect bounds was
completed in 1986, and we sought to update it and improve on it in various
aspects: First, we considered  all possible scalar leptoquarks while the work
in \cite{Buch1} dealt only with the superstring inspired $E_6$ leptoquark,
called $S$ in our paper. Second, there are new experimental results which
enable us to derive considerably stronger bounds. In particular there has been
a lot
of progress in both experimental measurements and theoretical calculations for
atomic parity violation and universality in leptonic $\pi$ decays. Third, we
take into account bounds from \KKb{} and \DDb{} mixing. The significance of
these bounds was pointed out only recently \cite{me}. Finally, we extract,
for each leptoquark, only the {\it utterly unavoidable} bound on its mass and
its coupling to the first generation. Obviously, these are the relevant bounds
for the direct searches in HERA, as well as for other direct and indirect
searches.

Our final bounds are presented in figure 1, where the mass range extends
to the multiTeV range. This figure can be used to examine the feasibility
of
methods proposed for leptoquark searches in various machines.  The bounds can
also be read from table 1, 2 and 7. The tables are convenient to use since they
give the lower bound on the mass as a simple function of the coupling constant,
but for some leptoquarks the bounds in the tables are somewhat weaker than the
full bounds presented in the figures.  In figure 2, we also compare our bounds
to the
first HERA results.

Since we are interested in the ``utterly unavoidable'' bounds, let us set the
stage for them by reviewing the means for circumventing other bounds.
Basically, there are three requirements that leptoquarks should obey in order
to evade some of the strongest bounds on their parameters: They should {\it
not} couple to diquarks, and they should couple chirally and diagonally. We
will now explain in some detail the meaning of these conditions:\newline
$\bullet$ Diquark couplings are forbidden since they, together with the
lepton-quark couplings lead to nucleon decay. The bound on the leptoquark mass
is then extremely strong, of the order of the scale of grand-unified theories.
\newline
$\bullet$ When we say that a leptoquark couples chirally, we mean that it
couples {\it either} to left--handed (LH) {\it or} to right--handed (RH)
quarks, but not to both. A nonchiral leptoquark induces the following
four-Fermi
interaction:
\beq
\L_{eff}=\frac{g_L g_R}{2M^2}~\bar u_R d_L~\bar e_R \nu_L
\label{gLgR}\;,
\eeq
where $M$ is the leptoquark mass and $g_L$ and $g_R$ are its couplings to LH
and RH quarks respectively. The above interaction contributes to
$\pi\longrightarrow e\nu$ decay and, in contrast to the standard model
interaction, it is not chiral and its amplitude is not helicity suppressed. The
amplitude is therefore enhanced by $m_\pi/m_e$ relative to the standard model
amplitude and, in addition, it is possible to show that there is further
enhancement by $m_\pi/(m_u+m_d)$ \cite{Shanker}. The enhanced effect of the
interaction (\ref{gLgR}) leads to unacceptable deviations from lepton
universality in $\pi$ decays, unless one strongly constrains the leptoquark
parameters with the 95\% CL bound as strong as $M^2/|g_L g_R|\geq (100~{\rm
TeV})^2$. The chirality requirement enables us to circumvent this bound.
\newline
$\bullet$ Leptoquarks couplings are called ``diagonal'' when the
leptoquark couples to a single leptonic generation and to a single quark
generation. If the leptoquarks couple nondiagonally they induce flavour
changing neutral current (FCNC) processes in both the leptonic sector and the
quark sector, leading to strict bounds on the leptoquark parameters
\cite{PatiS}, \cite{Buch1}. To avoid these bounds we impose diagonality of the
couplings.
However, we recently pointed out \cite{me} that diagonality is not really
possible for leptoquarks that couple to left-handed quarks. The fact that the
CKM matrix \cite{CKM} is not trivial implies that one cannot diagonalize the
leptoquark interactions simultaneously in the up and the down quark sectors.
For example, if the couplings to the up sector are diagonal, and the leptoquark
couples only to the first generation up quark, then the couplings in the down
sector are not diagonal: The leptoquark couples ``mainly'' to the down quark,
but there is also some coupling to the strange quark (suppressed by
$\sin\theta_C$) and some coupling to the bottom quark (suppressed by $V_{13}$,
where $V$ is the CKM matrix). Similarly, if the leptoquark couples diagonally
to the down quark, then its couplings to the up quark sector are {\it almost}
diagonal, but not strictly so. In the following, we assume approximate
diagonality of the leptoquark couplings to LH quarks: the leptoquarks couple
mainly to the first generation, with their couplings to the second and third
generations suppressed by $O(\sin\theta_C)$ and $O(|V_{13}|+|V_{12}V_{23}|)$,
respectively. Approximate diagonality softens the FCNC bounds, but does not
avoid
them completely. In section 6 we shall analyse this
problem in detail, and show that the FCNC bounds from the two sectors combine
to give a significant and unavoidable bound on the {\it flavour conserving}
coupling of the leptoquark to the first generation.

We should stress that the unavoidable bounds, which are the subject of this
paper, are {\it independent} of the above assumptions on the leptoquarks
couplings. These assumptions are just a matter of convenience: With them,
avoidable bounds are circumvented and the discussion of the unavoidable bounds
simplifies.

In addition to the assumptions on the leptoquark couplings, we make two
``working assumptions'': First, we assume that at most one leptoquark multiplet
exists. Second, we ignore mass splitting within a leptoquark multiplet. With
these assumptions the presentation of bounds simplifies considerably, as there
are only two parameters: a single coupling and a single mass.
In Appendix B
we discuss the modification of our bounds when the working assumptions are
dropped.

The rest of the paper is organized as follows: In the following section we
present the leptoquarks and their interactions and introduce notation, then we
turn to bounds: In section 3 we quote the bounds on the leptoquark parameters
from the direct searches at LEP, UA2 and CDF. Sections 4 to 6 discuss the
strongest indirect bounds we find: Section 4 deals with atomic parity
violation, Section 5 with universality in leptonic $\pi$ decays and section 6
with FCNC bounds: Section 6.1 is introductory, section 6.2 discusses rare
$K$ decay bounds, and section 6.3 describes neutral meson mixings bounds. In
section 6.4 we combine the FCNC bounds from the two quark sectors to a bound
on the flavour conserving coupling to the first generation. Section 7 is a
summary of our results.
We have relegated to Appendix A several bounds that are weaker than those of
sections 4 to 6.  These include bounds from $eD$
scattering, $p\bar p$ scattering to $e^+e^-$, hadronic forward--backward
asymmetry in $e^+e^-$ accelerators and universality in leptonic $K$ decays.
In Appendix B we consider the modification of our bounds when the ``working
assumptions'' are dropped.

\section{The scalar leptoquarks and their interactions}

The list of all possible scalar leptoquarks \cite{Buch2} includes the $S$ and
the $\tilde S$ leptoquarks in the $(0)_{1/3}$ and $(0)_{4/3}$ representations
of $SU(2)_{\rm W}\times U(1)_{\rm Y}$, the $D$ and $\tilde D$ leptoquarks in
the $(1/2)_{-7/6}$ and $(1/2)_{-1/6}$ representations, and the $T$ leptoquark
in the $(1)_{1/3}$ representation.

Some of these leptoquarks are forced to couple chirally by their \sutw{}
representations: $\tilde S$ and $\tilde D$ can couple only to RH quarks, $T$
only to LH quarks. The other leptoquarks, $S$ and $D$, can couple either to RH
or to LH quarks. We will call these leptoquarks $S_R$ and $D_R$ when they
couple to RH quarks and $S_L$ and $D_L$ when they couple to LH quarks. Note
that our subscripts $R$ and $L$ are determined by the quark helicities, in
contrast to the notation in \cite{Buch2}, which is fixed by the lepton
helicity. As a result, our notation for the subscript on the $D$ leptoquark is
opposite to the one of \cite{Buch2}.

The Yukawa interactions of the leptoquarks that couple to RH quarks
are given by:
\begin{eqnarray}
\L_{S_R} &=& g~\bar e^c u_R  \, S_{R}^{(1/3)}\nonumber\\
\L_{\tilde S} &=& g~\bar e^c d_R  \,\tilde S^{(4/3)}\nonumber\\
\L_{D_R} &=& g~\left(\bar e\, u_R D_R^{(-5/3)} +
                            \bar \nu\, u_R D_R^{(-2/3)} \right) \nonumber\\
\L_{\tilde D} &=& g\left(~\bar e\, d_R \tilde D^{(-2/3)} +
                            \bar \nu\, d_R \tilde D^{(1/3)} \right) \;,
\label{YukawaR}
\end{eqnarray}
where the superscripts on the leptoquark fields indicate their electromagnetic
charge.

The Yukawa couplings of the leptoquarks that couple to LH quarks are more
complicated. Here we need to introduce two sets of couplings: $g_i$ is the
coupling to the $i$'th up-quark generation, $g_i'$ is the coupling  to the
$i$'th down-quark generation and they are related by a CKM rotation:
$g_i'=g_j V_{ji}$, with $V$ the CKM mixing matrix\footnote{The normalization
of the couplings of the $T$ leptoquark is the one used
in \cite{Buch2}. The $T$ couplings we used in \cite{me} are larger by
$\sqrt2$ than the couplings we use here.}
\begin{eqnarray}
\L_{S_L} &=& \sum_i \, \left(g_i~\bar e^c u^i_L - g_i'~\bar\nu^c d^i_L \right)
               \, S_{L}^{(1/3)}\nonumber\\
\L_{D_L} &=& \sum_i \, \left\{g_i~\bar e\, u^i_L D_L^{(-5/3)} +
                            g_i'~\bar e\, d^i_L D_L^{(-2/3)} \right\}
\nonumber\\
\L_T &=& \sum_i \, \left\{{\sqrt2} g_i~\bar \nu^c u^i_L T^{(-2/3)}
      + (g_i~\bar e^c u^i_L + g_i'~\bar\nu^c d^i_L) \, T^{(1/3)}
      + {\sqrt2} g_i'~\bar e^c d^i_L T^{(4/3)} \right\}  \; .
\label{YukawaL}
\end{eqnarray}

In  order to present our bounds we define the overall strength of the Yukawa
couplings to be $g$, with
\beq
g=\sqrt{\sum_i~|g_i|^2}
\label{goverall}\;,
\eeq
and give our final results as bounds in the $g$ -- $M$ plane. Note that since
we assume that the leptoquarks couple mainly to the first generation, the
second and third generation couplings are suppressed by $O(\sin\theta_C)$ and
$O(|V_{13}|+|V_{12}\cdot V_{23}|)$. The first generation couplings are then
equal to $g$ to a very good approximation (up to $2-3\%$), and in the
following we will often ignore the differences between $g$, $g_1$ and $g_1'$.

For convenience, we also introduce the parameters $\eta_I$, with $I$ running
over all leptoquark multiplets: $I=S_L,S_R,\tilde S,D_L,D_R,\tilde D,T$.
$\eta_I$ gets the value $1$ when we consider a theory with the leptoquark $I$,
and otherwise it vanishes.

\section{Bounds from direct searches in LEP and TEVATRON}

The LEP experiments searched for leptoquark pair production  in $Z$ decays. No
evidence for such a decay mode was found and consequently LEP set a lower bound
on the leptoquark mass: $M\gtap M_Z/2$ \cite{LEPdirect}.

UA2 \cite{UA2} and CDF \cite{CDFdir} searched for leptoquark pairs produced via
an intermediate gluon. In contrast to LEP, where one can search for all types
of leptoquark pair events, namely (i) events with both leptoquarks decaying to
a charged lepton and a jet, (ii) events with one leptoquark decaying to a
charged lepton and a jet and the other to a neutrino and a jet, and (iii)
events with both leptoquarks decaying to a neutrino and a jet, the UA2
experiment did not search for the last type of events, and CDF did not search
for the last two types of events. Consequently,
the bounds from these experiments depend on  $b$, the branching ratio of the
decay of the  leptoquark to a charged lepton and a quark: If $b=1/2$ CDF bounds
the leptoquark mass to lie above $80$~GeV, and if $b=1$ to lie above $113$~GeV
\cite{CDFdir}. Studying the interactions (\ref{YukawaR}) and (\ref{YukawaL}),
one sees that $S_L$ has $b=1/2$ and its mass is therefore constrained to lie
above $80$~GeV. All the other leptoquark multiplets contain at least one
component with $b=1$. Under our working assumption of no mass splittings within
a leptoquark multiplet, we find that all the leptoquarks, but $S_L$, are
heavier than $113$~GeV.

\section{Atomic parity violation}

Measurements of atomic parity violation have not previously been used to
set bounds
on leptoquarks, although it was pointed out in ref.~\cite{Lang} that such
bounds could be very significant.  In fact, recent improvement on measurements
of atomic parity violation in Cesium as well as improved theoretical
calculations
turn out to lead to very strong bounds.
The relevant quantity is the Cesium ``weak charge''
defined by:
\beq
Q_W=-2\left[C_{1u}(2Z+N) + C_{1d}(2N+Z)\right]
\label{QW}\;,
\eeq
with $C_{1u}$ and $C_{1d}$
defined {\it e.g.} in \cite{PDG} and with $Z=55$ and $N\simeq77.9$ for Cesium.
The latest experimental result \cite{Csexp} and the standard model estimate
\cite{Cstheo} for $Q_W$ are:
\begin{eqnarray}
Q_W^{\rm exp}&=&-71.04\pm1.81 \nonumber\\
Q_W^{\rm SM}&=&-73.12\pm0.09
\label{qwexpsm}\;.
\end{eqnarray}
In a theory with a leptoquark, there is an additional contribution to $Q_W$,
given by:
\beq
\eqalign{
\Delta Q_W^{LQ}=-2\left(\frac {g/M}{g_W/M_W}\right)^2
\left[\vphantom{(Z+2N)\eta_{S_L}}\right.&(2Z+N)
\cdot(-\eta_{S_L}+\eta_{S_R}-\eta_{D_L}+\eta_{D_R}-\eta_T) \cr
             &\left. + (Z+2N)
\cdot(\eta_{\tilde S}-\eta_{D_L}+\eta_{\tilde D}-2\eta_T) \right]
\label{qwlq}
}
\eeq
Here $g$ and $M$ are the coupling and mass of the leptoquarks and $g_W$ and
$M_W$ are the coupling and mass of the $W$ boson.
The close agreement between the experimental $Q_W$ value and the standard model
estimate (see equation (\ref{qwexpsm})) leads to strong bounds on $g/M$.
These are summarized in table 1.

\begin{table}
\begin{center}
\begin{tabular}{|l|c|c|c|c|c|c|c|}\hline
&$S_L$&$S_R$&$\tilde S$&$D_L$&$D_R$&$\tilde D$&$T$\\ \hline
$M_{4\pi}$&
3600 & 7000 & 7400 & 5200 & 7000 & 7400 & 6400\\ \hline
$M_1$&
1000 & 2000 & 2100 & 1500 & 2000 & 2100 & 1800\\ \hline
$M_e$&
305 & 600 & 630 & 440 & 600 & 630 & 550\\ \hline
\end{tabular}
\medskip
\caption[table1]{
\it Atomic parity violation 95\% CL lower bounds on the ratio $M/g$,
in GeV. We present the bounds  in three equivalent ways in  order to simplify
the comparison to the various notations used in other leptoquark papers.
$M_{4\pi}$ is the lower bound on the leptoquark mass when the coupling becomes
nonperturbative $g^2=4\pi$ , $M_1$ is the bound when the coupling is 1 and it
is thus the bound on $M/g$ and $M_e$ is the bound when the coupling is equal to
the electromagnetic coupling $g=e$.}
\end{center}
\end{table}

The bounds we will discuss in the following sections apply only to the
leptoquarks that couple to LH quarks. Table 1 therefore contains our final
bounds on the leptoquarks that couple to RH quarks ($S_R$, $\tilde S$, $D_R$
and $\tilde D$) and these can be summarized by $M/g\gtap 2$~TeV.

\section{Bounds from universality in leptonic $\pi$ decays}

A remarkable progress has been achieved in both experimental and theoretical
research of leptonic $\pi$ decays. There have been two new experiments, one in
TRIUMF\cite{TRIUMPH}, the other in PSI \cite{PSI}. Combining their results we
find:
\beq
R^{\rm exp}=(1.2310\pm0.0037)\cdot 10^{-4}
\label{Rexp}
\eeq
where $R=BR(\pi\longrightarrow e\nu)/BR(\pi\longrightarrow \mu\nu)$.

The theoretical standard model calculation by Marciano and Sirlin has been
updated
\cite{MaSi}
and it now yields:
\beq
R^{SM}=(1.2352\pm0.0005)\cdot 10^{-4}
\label{Rthe}
\eeq
The theoretical prediction in a theory with a leptoquark is:
\beq
R^{LQ}=R^{SM}
\left(1+\left(\frac {g/M}{g_W/M_W}\right)^2\cdot(\eta_{S_L}-\eta_T)\right)^2
\label{Rlq}
\eeq
Equations (\ref{Rexp}-\ref{Rlq}) lead to the bounds of table 2. Note that for
$S_L$ the bound on $M/g$ of the leptoquark is considerably stronger than the
bound from atomic parity violation, while for the $T$ leptoquark the two bounds
(universality in leptonic $\pi$ decays and atomic parity violation) are
essentially equal.
\begin{table}
\begin{center}
\begin{tabular}{|l|c|c|}\hline
&$S_L$&$T$\\ \hline
$M_{4\pi}$&
12000 & 6400\\ \hline
$M_1$&
3400 & 1800\\ \hline
$M_e$&
1040 & 540\\ \hline
\end{tabular}
\medskip
\caption[table2]{\it 95\% CL bounds on the ratio $M/g$, in GeV, from
universality in
leptonic $\pi$ decays.}
\end{center}
\end{table}

\section{Bounds from FCNC processes}
\subsection{Introduction to FCNC bounds}

As mentioned above, leptoquarks that couple to LH quarks have two sets of
coupling constants, $g_i$ is the coupling to the up-like quark of the $i$th
generation and $g_i'$ are the couplings to the down-like quarks. The $g_i$ and
$g_i'$ are related through a CKM rotation. Since we consider leptoquarks that
couple mainly to the first generation, the third generation couplings are so
suppressed that they have actually no effect. We therefore ignore them and
reduce to a two generation picture, so that:
\begin{eqnarray}
g_1=g\cos\theta~~~~~&{\rm and}&~~~~~g_2=-g\sin\theta \nonumber \\
g'_1=g\cos(\theta_C-\theta)~~~~~&{\rm
and}&~~~~~g'_2=g\sin(\theta_C-\theta)
\label{gpar}\;.
\end{eqnarray}
The angle $\theta$ describes the deviation from diagonality in the up sector,
while $(\theta_C-\theta)$ describes the deviation from diagonality in the down
sector. $\theta$ therefore determines the division of the FCNC problems between
the two quark sectors. Note that we do not consider the possibility of a
nontrivial phase between $g_1$ and $g_2$. Such a phase leads to very severe
bounds, since the leptoquarks will contribute to the $\epsilon$ parameter of
\KKb{} mixing \cite{Shanker}. These bounds are stronger by
$\sqrt{(\sin2\alpha)/(2\sqrt{2}\epsilon)}$ than the \KKb{} mixing bounds of
table 5, where $\alpha$ is the phase. Since we are interested only in the {\it
unavoidable} bounds on the leptoquark couplings, we discard the case of complex
couplings.

The FCNC bounds from the up sector apply to the coupling
combination
$|g_1g_2|$ and the FCNC bounds from the down sector to the combination
$|g'_1g'_2|$. In the following sections we give the upper bounds on these
coupling constants combinations as a function of the leptoquark mass $M$, and
in section 6.4 we combine these bounds into bounds on $g$.

\subsection{Bounds from $K$ decays}

Leptoquarks induce the rare $K$ decays \Kpinunu{} and \Kee{}.
\Kpinunu{} decay is induced by the $S_L$ and $T$
leptoquarks via the effective interaction:
\beq
\L_{eff}= \frac{g_1'{g_2'}}{2M^2}~\bar s\gamma_\mu P_L d
                                 ~\bar \nu\gamma^\mu P_L \nu
                                 ~(\eta_{S_L}+\eta_T)
\label{LKpinunu}\;,
\eeq
where $P_L=(1-\gamma_5)/2$ is the LH projection operator. The 95\% CL
experimental bound on the \Kpinunu{} decay rate \cite{Atiya} is
\beq
{\rm BR}(K^+\longrightarrow\pi^+\nu\bar\nu)\leq 6.8\cdot 10^{-9}\;.
\label{BRKpinunu}
\eeq
Comparing the branching ratio induced by eq.~(\ref{LKpinunu}) to
eq.~(\ref{BRKpinunu}) leads to the bounds of table 3.

\begin{table}
\begin{center}
\begin{tabular}{|l|c|c|}\hline
&$S_L$&$T$\\ \hline
$|g_1'g_2'|\leq$ & $1.86\cdot10^{-8}\sin\theta_C~M^2$ &
$1.86\cdot10^{-8}\sin\theta_C~M^2$
\\ \hline
\end{tabular}
\medskip
\caption[table3]{\it \Kpinunu{} decay 95\% CL bounds on the coupling constant
combination $g_1'g_2'$. The bounds are given as a function of the leptoquark
mass $M$, with $M$ in GeV.}
\end{center}
\end{table}
\begin{table}
\begin{center}
\begin{tabular}{|l|c|c|}\hline
&$D_L$&$T$\\ \hline
$g_1'g_2'\leq$&
$2.92\cdot10^{-8}\sin\theta_C~M^2$&$1.46\cdot10^{-8}\sin\theta_C~M^2$
\\ \hline
\end{tabular}
\medskip
\caption[table4]{\it \Kee{} decay 95\% CL bounds on the coupling constant
combination $g_1'g_2'$. The bounds are given as a function
of the leptoquark mass $M$, with $M$ in GeV.}
\end{center}
\end{table}

\Kee{} decay is induced by the $D_L$ and $T$ leptoquarks via the
effective interaction:
\beq
\L_{eff}= \frac{g_1'{g_2'}}{2M^2}~\bar s\gamma_\mu P_L d
                                 ~\left(2\eta_T\bar e\gamma^\mu P_L e
                                 - \eta_{D_L} \bar e\gamma^\mu P_R e\right)
\label{LKee}\;,
\eeq
where $P_L$ and $P_R$ are the LH and RH projection operators, respectively.
The 95\% CL experimental bound on the \Kee{} decay rate \cite{Ari} is
\beq
{\rm BR}(K_L\longrightarrow e^+e^-)\leq 5.3\cdot 10^{-11}\;.
\label{BRKee}
\eeq
Comparing the branching ratio induced by eq.~(\ref{LKee}) to
eq.~(\ref{BRKee}) leads to the bounds of table 4.

\subsection{Bounds from neutral meson mixings}

The $S_L$, $D_L$ and $T$ leptoquarks induce new contributions to \KKb{} and
\DDb{} mixing via loops of leptons and leptoquarks. One could, at first
thought, discard the bounds from neutral meson mixings as unimportant, since
they arise only at one loop, in contrast to other leptoquark bounds that arise
already at tree level. However, such an approach is mistaken: After all, \KKb{}
and \DDb{} mixing arise in the standard model too only at one loop. Moreover,
the GIM mechanism of the standard model leads to a suppression of {\it e.g.}
\KKb{} mixing by $(m_c/M_W)^2$, while for the leptoquarks contribution there is
no suppression of this kind. We therefore should expect neutral meson
mixing to give us significant bounds on the leptoquarks parameters.

The leptoquarks contribution to \KKb{} and \DDb{} mixing are given by:
\begin{eqnarray}
\Delta M_{12}^K&=&\frac{1}{192\pi^2M^2} (g_1'{g_2'})^2  f_K^2 B_K M_K
\cdot(\eta_{S_L}+\eta_{D_L}+5\eta_T)
\nonumber\\
\Delta M_{12}^D&=&\frac{1}{192\pi^2M^2} (g_1{g_2})^2  f_D^2 B_D M_D
\cdot(\eta_{S_L}+\eta_{D_L}+5\eta_T)
\label{mix}\;.
\end{eqnarray}
Demanding that the leptoquark contribution to \KKb{} mixing does not exceed the
measured value of
$\Delta M_{12}^K = 3.52\cdot 10^{-6}$~eV \cite{PDG},
and that the leptoquark contribution to \DDb{} mixing does not
exceed the 95\% CL experimental bound $\Delta M_{12}^D\leq 1.5\cdot 10^{-4}$~eV
\cite{FNALTPS} we are led to the bounds of tables 5 and 6. The values of the
$B$ parameters we used are $B_K=0.7$~ \cite{lattice} and  $B_D=1.0$, and for
the $D$ decay constant we took $f_D=0.25$~GeV.

\begin{table}
\begin{center}
\begin{tabular}{|l|c|c|c|}\hline
&$S_L$&$D_L$&$T$\\ \hline
$|g_1'g_2'|\leq$&
$1.25\cdot10^{-4}\sin\theta_C~M$&
$1.25\cdot10^{-4}\sin\theta_C~M$&
$5.58\cdot10^{-5}\sin\theta_C~M$
\\ \hline
\end{tabular}
\medskip
\caption[table5]{\it \KKb{} mixing bounds on the coupling constant
combination $g_1'g_2'$. The bounds are given as a function
of the leptoquark mass $M$, with $M$ in GeV.}
\end{center}
\end{table}

\begin{table}
\begin{center}
\begin{tabular}{|l|c|c|c|}\hline
&$S_L$&$D_L$&$T$\\ \hline
$|g_1g_2|\leq$&
$2.24\cdot10^{-4}\sin\theta_C~M$&
$2.24\cdot10^{-4}\sin\theta_C~M$&
$1.00\cdot10^{-4}\sin\theta_C~M$
\\ \hline
\end{tabular}
\medskip
\caption[table6]{\it \DDb{} mixing 95\% CL bounds on the coupling constant
combination $g_1g_2$. The bounds are given as a function
of the leptoquark mass $M$, with $M$ in GeV.}
\end{center}
\end{table}

Note that the \KKb{} and \DDb{} mixing bounds are different from all previous
bounds: The bounds from atomic parity violation, universality in leptonic $\pi$
decays and rare $K$ decays all apply to $g/M$ or
$g_1'g_2'/M^2$, so $g \propto M$. In contrast, the neutral
meson mixing bounds apply to $g_1'g_2'/M$ and $g_1g_2/M$, so $g \propto
\sqrt{M}$.
This difference is due to the fact that all previous
bounds arise from tree level leptoquark contributions, while \KKb{} and \DDb{}
mixing arise at the one loop level, and this turns out to be advantageous: The
bounds from neutral meson mixings, because of their different functional
dependence on the couplings and mass, {\it always} become the dominant bounds
at the high mass region.

\subsection{Combining the FCNC bounds to a bound on $g$}

In this section we will combine the FCNC bounds from the two quark sectors
to an {\it unavoidable bound} on the overall coupling $g$. Since $g$ is equal
to a very good approximation to $g_1$ and $g_1'$, this means that the FCNC
bounds combine to a bound on the {\it flavour conserving} coupling of the
leptoquarks to the first generation.

We summarize the FCNC bounds in the following manner:
\begin{eqnarray}
f_u(M)&\geq|g_1g_2|&=g^2\sin(2|\theta|)/2 \nonumber\\
f_d(M)&\geq|g_1'g_2'|& =g^2\sin(2|\theta_C-\theta|)/2
\label{gcombi}\;,
\end{eqnarray}
where $f_u(M)$ and $f_d(M)$ are the strongest FCNC bounds of the up and down
quark sectors respectively, and can be read from tables 3, 4, 5 and 6.
Equations (\ref{gcombi}) make it clear that {\it any} angle $\theta$
leads to bounds on $g^2$. We are interested in the unavoidable bound on the
coupling and we therefore look for the ``best'' angle $\theta$, \ie{}
the one that
leads to the softest bounds on $g^2$. This angle is given by simultaneously
saturating the two inequalities in (\ref{gcombi}), so that:
\beq
\frac{f_u(M)}{f_d(M)}=\left|\frac{\sin2\theta}{\sin2(\theta_C-\theta)}\right|
\label{theta}\;.
\eeq
Solving equation (\ref{theta}) for the ``best'' angle $\theta$,
\beq
\tan(2\theta^{\rm best})=\frac{\sin2\theta_C}{f_d/f_u+\cos2\theta_C}
\label{best}\;,
\eeq
and substituting this angle into either of the two inequalities of
(\ref{gcombi}),
we get the {\it unavoidable} FCNC bound on the overall coupling $g$:
\beq
g^2(M)\leq 2f_u(M)/\sin2(\theta^{\rm best}(M))
\label{FCNCbound}\;.
\eeq

Again, we wish to stress \cite{me} that the FCNC bound always become the
most stringent bound in the high mass region. To see that,
note that in this region both $f_u$ and $f_d$ are linear in the leptoquark
mass: $f_u$ is the \DDb{} mixing bound, and is therefore always linear in $M$.
$f_d$ is the strongest of the rare $K$ decay bounds and the \KKb{} mixing
bound. Since the rare $K$ decays bounds on $g_1'g_2'$ are quadratic in $M$
while the \KKb{} mixing bound is linear, the latter will dominate at high
masses. Therefore, at high masses, the ratio $f_d/f_u$ is independent of the
leptoquark mass; consequently the ``best'' angle $\theta$ is also $M$
independent (see equation \ref{best}), and the bound on $g^2$ is linear in $M$
(see equation \ref{FCNCbound}). In contrast, the atomic parity violation and
universality in leptonic $\pi$ decay bounds on $g^2$ are quadratic in $M$. The
combined FCNC bound will therefore always dominate at high enough masses.
Indeed, we find that the FCNC bound dominates above 3600~GeV in the case of
$S_L$, but already above 570~GeV and 390~GeV in the cases of $D_L$ and $T$
respectively.

In table 7 we list the combined FCNC bound from \DDb{} and \KKb{} mixing.
This is a true FCNC bound, although in the low mass region there are stronger
FCNC bounds combined from \DDb{} mixing and rare $K$ decays.

\begin{table}
\begin{center}
\begin{tabular}{|l|c|c|c|}\hline
&$S_L$&$D_L$&$T$\\ \hline
$M_{4\pi}$&
35,500&35,500&79,500
\\ \hline
$M_1$&
2,800&2,800&6,300
\\ \hline
$M_e$&
260&260&580
\\ \hline
\end{tabular}
\medskip
\caption[table7]{\it
Combined \KKb{} and \DDb{} mixing lower bounds on $M/g^2$ at 95\% CL, in GeV.
$M_{4\pi}$ and $M_e$ are again the lower bounds on the mass when the coupling
constant is set to $g^2=4\pi$ and $e^2$, respectively. $M_1$ is the bound on
the mass when the coupling constant is set to 1, and it is therefore
also the bound on
$M/g^2$. Note the different functional dependence on the coupling constant
relative to tables 1 and 2.}
\end{center}
\end{table}

\section{Summary}
We made a comprehensive survey of the bounds on scalar leptoquarks couplings to
the first generation. We have discarded bounds that can be avoided, and
concentrated
only on those bounds that are completely inescapable. We found that the most
stringent bounds arise from low energy data: Atomic parity violation,
universality in leptonic $\pi$ decays and FCNC processes: \Kpinunu{} decay,
\Kee{} decay and \KKb{} and \DDb{} mixing.

\psinsert{.9}{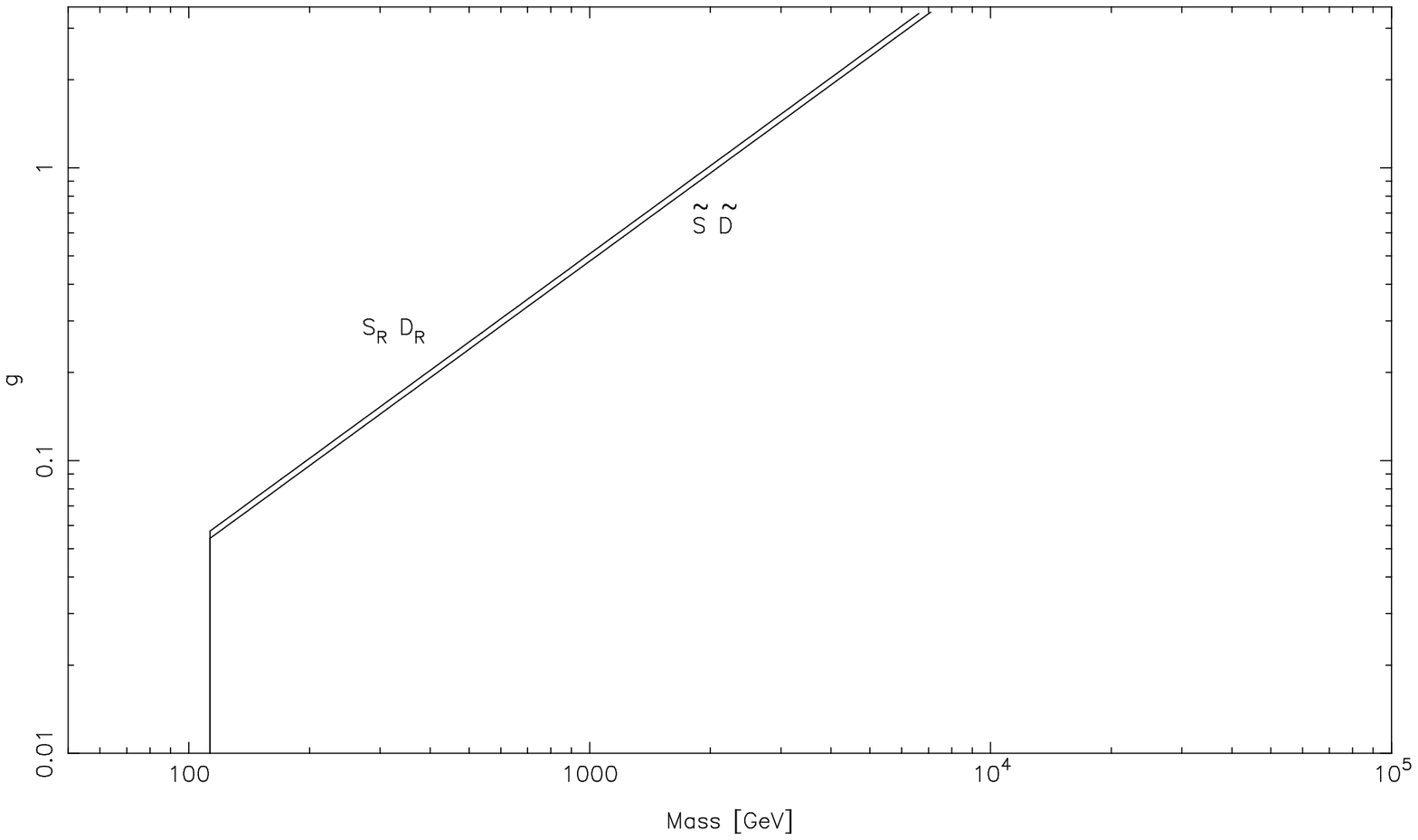}{{\bf Figure 1a.} The overall bound on leptoquarks
that couple to RH quarks.  The regions above the lines are excluded.
The graph is cut off at $g^2=4\pi$.}

\psinsert{.9}{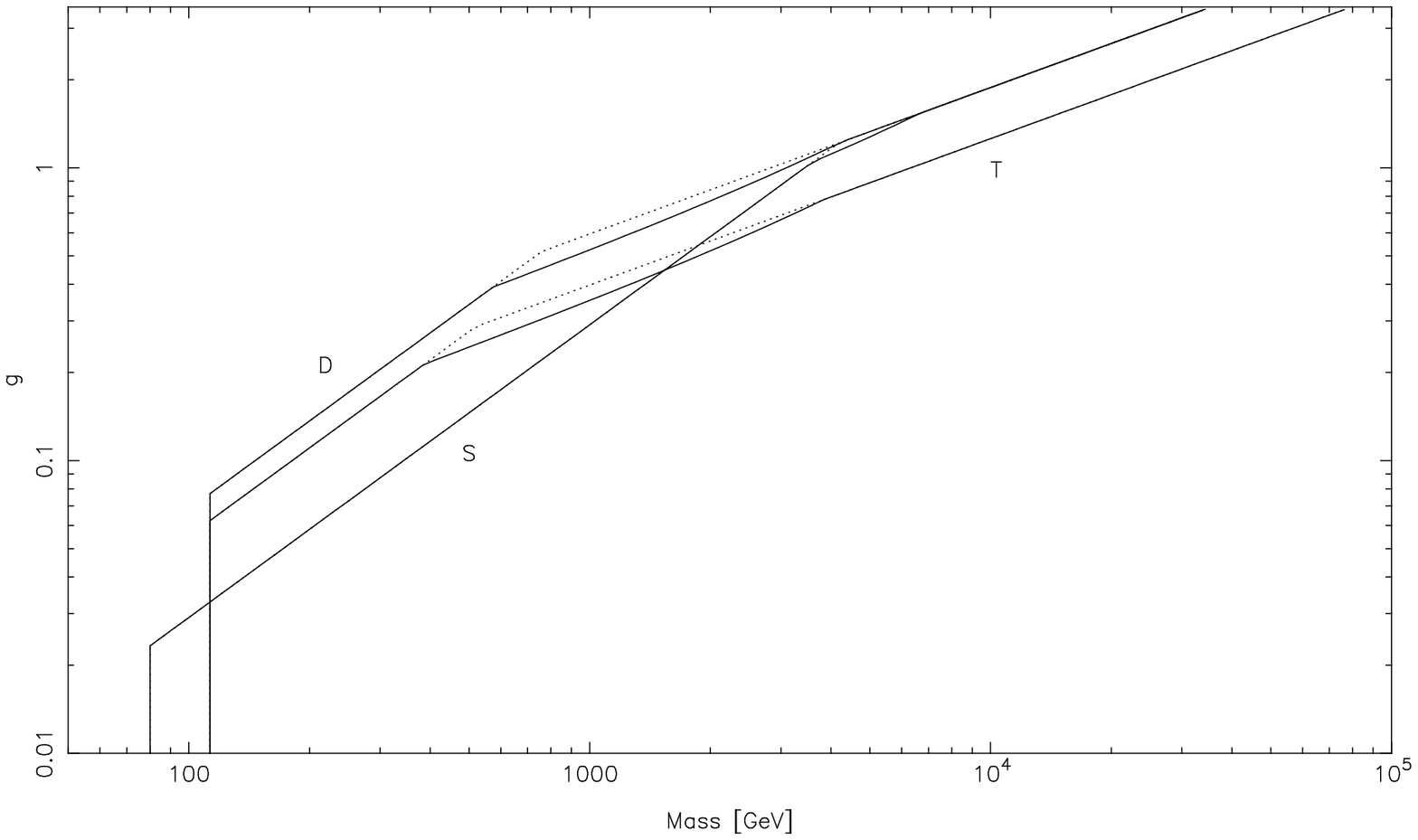}{{\bf Figure 1b.}
The overall bound on leptoquarks that couple to LH quarks.  The regions above
the lines are excluded.   The graph is cut off at $g^2=4\pi$.
The full lines show the exact bounds, the dotted lines
the approximate bounds of tables 1, 2 and 7.}

Our final bounds can be summarized in a few different ways: Figures 1a and 1b
show the overall bound on $g$ as a function of $M$ for all the leptoquarks.
Figure 1a describes the bounds on the leptoquarks that couple to RH quarks;
these come from atomic parity violation and are also given in table~1. Figure
1b describes the bounds on the leptoquarks that couple to the LH quarks, and
here one distinguishes three mass regions for each of the leptoquarks:
In the low mass region the dominant bound arises from atomic parity violation
or from universality in leptonic $\pi$ decays and it depends on $g/M$. In the
high mass region the most stringent bound is the FCNC bound derived by
combining the \KKb{} and \DDb{} mixing bounds, and it depends on $g^2/M$.
There is also an intermediate mass region, where the strongest
bound is the FCNC bound combined from rare $K$ decays and \DDb{} mixing. The
functional dependence of this bound on $g$ and $M$ is more complicated. Note
that the FCNC bounds exclude large new regions in the leptoquark parameter
space, and for $D_L$ and $T$ these bounds become dominant already at 570~GeV
and 390~GeV respectively. Figure 1b also contains the approximate bounds one
would get when ignoring rare $K$ decays. In this case, there are only two mass
regions for each leptoquark -- at low masses the bound depends on $g/M$, at
high masses on $g^2/M$. The approximate bounds can also be read from tables 1,
2 and 7; they have the advantages of being true bounds, being relatively good
approximations (the difference between the approximate and exact bounds on $g$
is at most 15\% for all masses) and most important, having simple functional
dependence on the leptoquark parameters. Figures 1a and 1b can be used to
estimate the feasibility of various methods proposed for leptoquark searches
\cite{proposals}. Our bounds already exclude large regions in the parameter
space that could be penetrated by some of these methods.

\psinsert{.9}{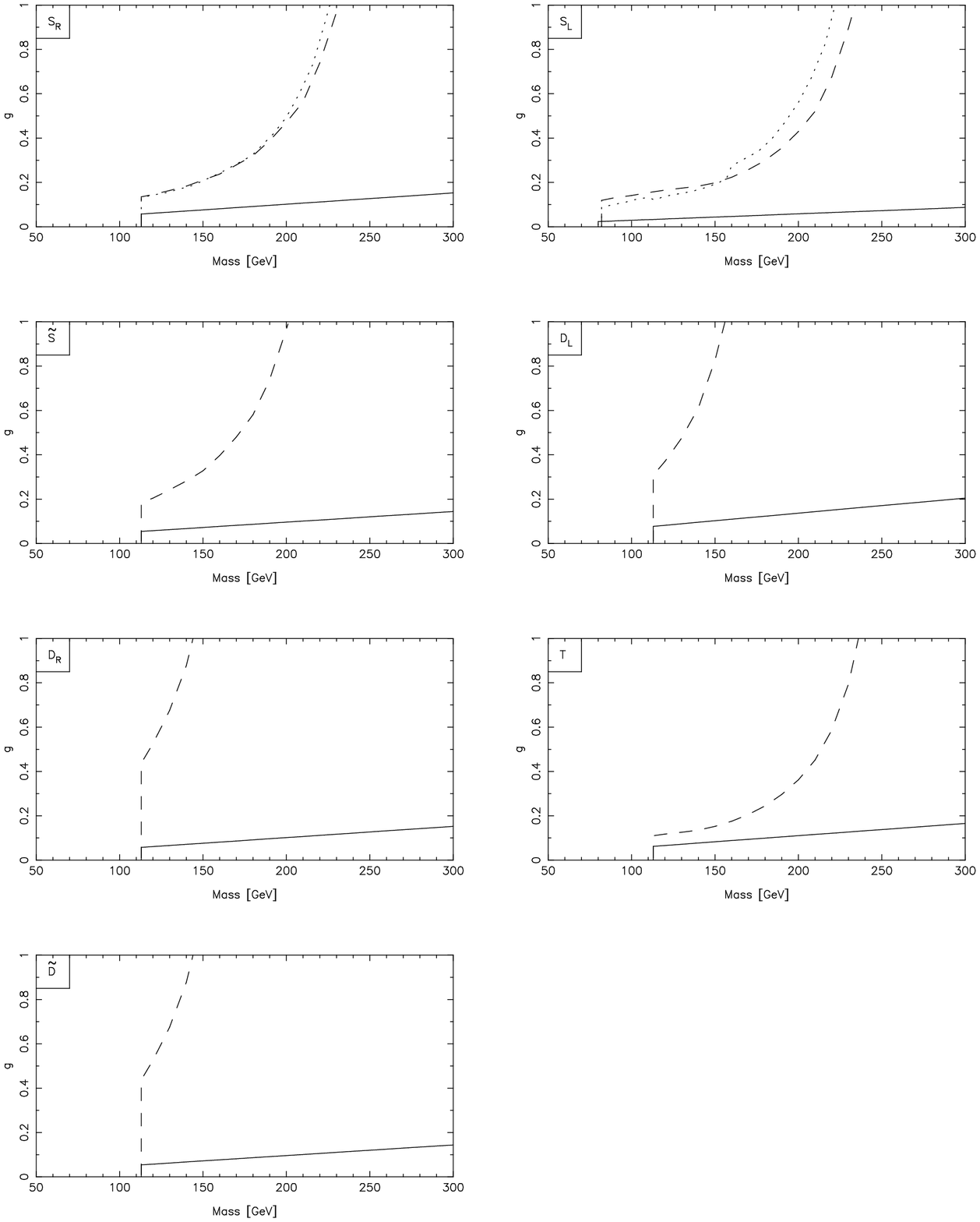}{{\bf Figure 2.} Direct and indirect bounds on
leptoquarks.  The solid lines are our bounds,
the dashed lines are the first bounds from the H1 group at HERA
and
the dotted lines the first bounds from the ZEUS group at HERA.}

In Figure 2 we restrict ourselves to the mass region which is subject to the
direct
searches at HERA. In this region our bounds on $g$ are linear in $M$ and can be
read off tables 1 and 2.
The figure compares our bounds to the first HERA results
\cite{HERA}, and one sees that at the moment our bounds are far stronger than
HERA's. In the future the situation will change, and HERA bounds {\it in this
mass region} will become far stronger than ours.
\begin{table}
\begin{center}
\begin{tabular}{|l|c|c|c|c|c|c|c|}\hline
&$S_L$&$S_R$&$\tilde S$&$D_L$&$D_R$&$\tilde D$&$T$\\ \hline
$M\geq$&1040&600&630&440&600&630&750
\\ \hline
\end{tabular}
\medskip
\caption[table8]{\it Final
upper bounds on the leptoquark masses in GeV, at 95\% CL,  when the coupling is
equal to the electromagnetic coupling, $g=e$.}
\end{center}
\end{table}

Finally, in table 8 we give the lower bound on the mass of the leptoquarks
when the
coupling constant is equal to the electromagnetic coupling $e$.

{\bf Acknowledgement}

I am indebted to Neil Marcus for many stimulating and fruitful discussions, as
well as to Ilan Levine, who have helped with some subtleties. I am grateful to
Kazuo Abe and Ichiro Adachi who provided me with the detailed data of their
respective groups, AMY and TOPAZ, and to Shmuel Nussinov, Avraham Montag and
Yossi Nir for helpful discussions.
Finally, I wish to thank Sacha Davidson, who pointed out an error in my
expression for $C_{1d}$, Gordy Kane who updated me on the results of the direct
leptoquark searches in CDF, and Laurence Littenberg and Doug Bryman who updated
me on the recent bounds on rare $K$ decays.

{\bf Note added in Proof:}
After this work was submitted for publication, we learned about another
recently completed research on leptoquark bounds by S.~Davidson, D.~Bailey
and B.A.~Campbell, Berkeley preprint CfPA 93--th--29, hep-ph/9309310.

\newpage

\appendix{Additional bounds}
In this appendix we present bounds from $eD$ scattering, $p\bar
p\longrightarrow e^+e^-$ scattering, hadronic forward--backward asymmetry in
$e^+e^-$ machines and universality in leptonic $K$ decays. All these bounds are
weaker than the ones in the body of the paper, but it is possible that in the
future better experimental data and improved theoretical estimates will enable
one to derive significant bounds from some of the processes discussed here.
Also, we should note that the bounds we get from hadronic forward-backward
asymmetry in $e^+e^-$ scattering apply to leptoquarks couplings to the electron
and the {\it first or second} generation of quarks, and for $\tilde S$ and
$\tilde D$ they apply to the couplings to the electron and any quark.

\subsection{$eD$ scattering}
$eD$ scattering provides information on the parity violating quantity
$C_{2u}-C_{2d}/2$ (for the definition of the $C_{2i}$ and their standard model
values see \cite{PDG}). The experimental result \cite{SLACeD} and the standard
model predictions are:
\begin{eqnarray}
(C_{2u}-C_{2d}/2)^{\rm exp}&=&-0.03\pm0.13\nonumber\\
(C_{2u}-C_{2d}/2)^{\rm SM}&=&-0.047\pm0.005
\label{eDexpsm}\;.
\end{eqnarray}
The additional contribution of a leptoquark is
\beq
\Delta(C_{2u}-C_{2d}/2)^{\rm LQ} = \left(\frac {g/M}{g_W/M_W}\right)^2
\cdot({-\eta_{S_L}+\eta_{S_R}-\eta_{\tilde
S}/2 +\eta_{D_L}/2-\eta_{D_R}+\eta_{\tilde D}/2})\;.
\label{eDlq}
\eeq
The agreement between the experimental result and the standard model prediction
leads to the bounds in table 9. These are considerably weaker than
the bounds derived from atomic parity violation and universality in leptonic
$\pi$ decays.
\begin{table}
\begin{center}
\begin{tabular}
{|l|c|c|c|c|c|c|c|}\hline
&$S_L$&$S_R$&$\tilde S$&$D_L$&$D_R$&$\tilde D$&$T$\\ \hline
$M_{4\pi}$&
910 & 810 & 640 & 570 & 910 & 570 & $-$ \\ \hline
$M_1$&
260 & 230 & 180 & 160 & 260 & 160 & $-$ \\ \hline
$M_e$&
80 & 70 & 50 & 50 & 80 & 50 & $-$ \\ \hline
\end{tabular}
\medskip
\caption[table9]{\it $eD$ scattering 95\% CL bounds on $M/g$, in GeV.}
\end{center}
\end{table}

\subsection{$p\bar p$ scattering to $e^+e^-$}
$p\bar p$ scattering to $e^+e^-$ was studied by the CDF group
\cite{CDF}. Analysis of the $e^+e^-$ mass distribution led to bounds on the
compositeness scales $\Lambda_{LL}^{-}\geq 2.2~$TeV and $\Lambda_{LL}^{+}\geq
1.7~$TeV  (for the definition of these scales see \cite{ELP}). We did not make
a detailed analysis, but estimate that similar bounds should apply to the
leptoquarks, namely, we expect bounds of the order of $M/g\geq 2~{\rm
TeV}/\sqrt{4\pi}$. These bounds are also weaker than
the bounds in the body of the paper.
Our conclusion is therefore that at present $p\bar
p\longrightarrow e^+e^-$ scattering does not provide useful bounds. We
do however recommend that future analysis of this process be used for
deriving bounds on leptoquarks since with improved statistics this may lead to
interesting results.

\subsection{Hadronic forward--backward asymmetry in $e^+e^-$ colliders}

To avoid possible confusion, we first comment
on an earlier work on a similar subject \cite{HeRi}. The authors of \cite{HeRi}
studied the scattering processes $e^+e^-\longrightarrow c\bar c$ and
$e^+e^-\longrightarrow b\bar b$ at $\sqrt{s}=40~$GeV, and derived bounds on
leptoquarks by requiring that the total cross section and the forward--backward
asymmetry for both these processes deviate at most by a few percent from the
standard model prediction. Although these are interesting bounds they are of
{\it no relevance} to our study: The bounds of \cite{HeRi} apply to leptoquarks
that couple to quarks of the second and third generation while we are
interested in leptoquarks that couple to the first generation.

The relevant process for leptoquarks that couple to the first generation is
$e^+e^-\longrightarrow q\bar q$. Here  a particular scattering is called
``forward'' if the negatively charged quark or antiquark scatters into the
forward hemisphere of the electron beam. The hadronic forward--backward
asymmetry of this process was studied at PEP \cite{PEP}, in PETRA \cite{JADE},
in TRISTAN \cite{TRISTAN} and in LEP \cite{LEP}. We chose to concentrate on
the results of TRISTAN and LEP.
Considering LEP, we have concentrated on OPAL measurements of the
forward--backward asymmetry as these led to a somewhat more accurate
determination of $\sin^2\theta_W$. We used the OPAL value $\sin^2 \theta_W =
.2321\pm0.0033$ to constrain leptoquarks in the following way: We
calculated the forward--backward asymmetry in the standard model with the
central OPAL value for $\sin^2\theta_W$. Then we defined ``the 95\% CL
deviations'' by repeating the calculation with $\sin^2\theta_W$ removed by $\pm
1.96\sigma$ from the central value. Finally, we calculated the asymmetry with
$\sin^2\theta_W$ at its central value but with leptoquarks, and required that
the deviation from the standard model prediction did not exceed ``the 95\% CL
deviations''.   This gives $M/g\geq60-80~$GeV, the exact value
depending on the leptoquark type. These bounds are far weaker than the bounds
derived from atomic parity violation and universality in leptonic $\pi$ decays.

Forward-backward asymmetry in TRISTAN leads to more
interesting bounds on leptoquark parameters.  Two groups,
TOPAZ and AMY, have provided us with detailed data on their differential cross
sections.  Following the
procedure used by TOPAZ to set bounds on the compositeness scale, we derived
bounds on the leptoquarks parameters by comparing the experimentally measured
differential cross section to the prediction of the leptoquark theory. Our
results are summarized in table 10.
\begin{table}
\begin{center}
\begin{tabular}
{|l|c|c|c|c|c|c|c|}\hline
&$S_L$&$S_R$&$\tilde S$&$D_L$&$D_R$&$\tilde D$&$T$\\ \hline
$M_{4\pi}$&
530 & 1000 & 890 & 1800 & 1300 & 480 & 690 \\ \hline
$M_1$&
150 & 290 & 250 & 510 & 370 & 140 & 200 \\ \hline
$M_e$&
45 & 90 & 75 & 150 & 110 & 40 & 60 \\ \hline
\end{tabular}
\medskip
\caption[table10]{\it The 95\% CL lower bounds on $M/g$, in GeV, as
derived from TRISTAN data. We also find a small allowed region for the $S_R$
leptoquark for $M/g$ between $\sim 120$~GeV and $\sim 140$~GeV.}
\end{center}
\end{table}
Although these bounds are considerably weaker than the atomic parity violation
and universality in leptonic $\pi$ decay
bounds we find them interesting since they apply to any leptoquark that
couples chirally to the electron and to the first and/or the  second quark
generations. For the $\tilde S$ and $\tilde D$ leptoquarks these bounds apply
also when they couple to the $b$ quark of the third generation.

We should note that the bounds derived from TRISTAN apply to the
quantity $M/g$ only at the high mass region, where the leptoquark propagator
can
be approximated as $1/M^2$. At lower masses, propagator effects make it
impossible to describe the exact bound in terms of a simple function of $M$ and
$g$. However, the bounds on $M/g$ which are described in table 10
still apply to a good approximation: There is only $\sim 3\%$ correction
when $M=200~$GeV, $\sim 10\%$ correction when $M=113~$GeV and $\sim 18\%$
correction when $M=80~$ GeV, relative to the bounds in the table. All the
correction {\it weaken} the bound.  This weakening is because here the
leptoquark runs in the $t$ or $u$ channel, with a propagator
$1/(M^2-t)$ or $1/(M^2-u)$, which
is suppressed relative to $1/M^2$.

\subsection{Bounds from universality in leptonic $K$ decays}
Leptoquarks lead to deviations from universality in leptonic $K$
decays. This leads to bounds on $g_1g_2'$, which is equal, to a very good
approximation, to $g_1'g_2'$. Universality in leptonic $K$ decay therefore
bounds the same coupling constant combination as do FCNC processes in the down
sector.

Defining $R_K$ to be the ratio of the decay rates of \Kenu{} and
\Kmunu, we quote the observed ratio \cite{PDG} and the standard
model prediction (at tree level):
\begin{eqnarray}
R^{\rm exp}_{K}&=&(2.45\pm0.11)\cdot 10^{-5}\nonumber\\
R^{\rm SM}_K&=&(\frac{m_e}{m_\mu})^2
(\frac{M_K-m_e}{M_K-m_\mu})^2=2.57\cdot10^{-5}
\label{RKSM}\;.
\end{eqnarray}
Leptoquarks modify the theoretical prediction for $R_K$ to:
\beq
R^{LQ}_{K}=R^{SM}_K\left[1+2\frac{g_1{g_2'}}{g_W^2
\sin\theta_C\cos\theta_C}
(\frac{M_W}{M})^2\cdot(\eta_{S_L}-\eta_T)\right]
\label{RKLQ}\;.
\eeq
The agreement between the experimental result and the standard model value
(equations (\ref{RKSM})) lead to the bounds of table 11. These bounds are
considerably weaker than the rare $K$ decay bounds of section 6.2.
\begin{table}
\begin{center}
\begin{tabular}{|l|c|c|}\hline
&$S_L$&$T$\\ \hline
$|g_1g_2'|\leq$ & $4.2\cdot10^{-6}\sin\theta_C~M^2$ &
$4.2\cdot10^{-6}\sin\theta_C~M^2$
\\ \hline
\end{tabular}
\medskip
\caption[table11]{\it Universality in leptonic $K$ decay 95\% CL lower bounds
on the coupling constant
combination $g_1g_2'$. The bounds are given as a function of the leptoquark
mass $M$, with $M$ in GeV.}
\end{center}
\end{table}

\appendix{Comments on the ``working assumptions''}
In this appendix we will comment on our ``working assumptions''; the
assumption that there is at most one leptoquark multiplet, and the assumption
that there is no mass splitting within a multiplet.

At the mass region above $\sim1~$TeV, we expect that our bounds still hold:
Here electroweak breaking effects should be small: Mass splitting within a
multiplet should be small relative to the average mass, since otherwise the
$\rho$ parameter gets unacceptably large contributions. Mixings amongst the
multiplets can also be ignored when considering the processes discussed above.
Then, since we do not expect exact or almost exact cancellations among the
contributions of the various leptoquark multiplets, all our bounds should still
hold.

At low masses one cannot ignore electroweak breaking. The parameter space then
includes many mass parameters, as mass splitting within a multiplet as well as
mixing become significant. It is hard to extract a clear picture in the general
case, but it is possible to do so if we keep the assumption of a single
leptoquark multiplet: First we note that the $S_L$, $S_R$ and $\tilde S$
leptoquarks contain one component each, and so the second assumption of no mass
splitting is trivially true in their case. Therefore all the bounds derived
above still apply for these leptoquarks. With regard to the \sutw{} doublets
and the triplet: The direct CDF bounds as well as the bounds from atomic parity
violation and universality in leptonic $\pi$ decays still apply, with some
modifications, to the components that couple to the electron: For $D_R$ and
$\tilde D$, the direct CDF bound ($M\geq 113~$GeV) and the atomic parity
violation bound (table 1) still apply to the $D^{(-5/3)}$ and $\tilde
D^{(-2/3)}$) component. For $D_L$, the CDF bound still applies to both
components, and so does the atomic parity bound (see table 1) except the last
is weakened by $\sim\sqrt2$. For the $T$ multiplet, the case of $T^{(1/3)}$ and
$T^{(4/3)}$ are different: For $T^{(1/3)}$ the direct CDF bound is weakened and
it now reads $M\geq 80~$GeV; the bound from universality in leptonic $\pi$
decays still holds. For $T^{(4/3)}$ the direct CDF bound still applies without
modification $M\geq 113~$GeV, but the bound from atomic parity violation is
weakened by $\sqrt{2(Z+2N)/[2(Z+2N)+(2Z+N)]}\sim0.83$. The direct searches in
CDF, atomic parity violation and universality in leptonic $\pi$ decays
therefore still supply us with significant bounds on the leptoquark multiplet
components that couple to the electron. These are also the components that can
be searched for in HERA.

\end{document}